\documentstyle[prl,eqsecnum,aps]{revtex}

\begin{document}
\author{Jian Qi Shen \footnote{E-mail address: jqshen@coer.zju.edu.cn}}
\address{1. Center for Optical and Electromagnetic Research, State Key Laboratory of Modern Optical
Instrumentation, College of Information Science and Engineering;
  2. Zhejiang Institute of Modern Physics and Department of
  Physics, \\
   Zhejiang University, Hangzhou 310027, P. R. China}
\date{\today }
\title{Exact Solutions and Geometric Phases of Dipole Oscillator \\in Rapidly Varying Electric
Field \footnote{The problem in this paper is trivial, so that it
won't be submitted elsewhere for publication. However, two useful
approaches to the explicit solutions of the time-dependent
Schr\"{o}dinger equation are introduced in the present paper. }}
\maketitle

\begin{abstract}
The present letter obtains the exact solution and geometric phase
of the time-dependent Schr\"{o}dinger equation governing the
dipole oscillator in the exterior electric field, by making use of
the Lewis-Riesenfeld invariant theory and the invariant-related
unitary transformation formulation. It is shown that the geometric
phase presents the global property of the time evolution of the
dipole oscillator in time-dependent environments.

\end{abstract}

Many effects or properties such as the time evolution, geometric
phase and topological property of the time-dependent $L-S$ coupled
system and many-spin system (the Heisenberg spin system) as well
as the spinning magnetic moment in exterior magnetic fields have
been extensively studied by many
authors\cite{Bouchiat,Datta,Mizrahi,Gao0,Cen,Yan}. Yan {\it et
al.} discussed the time evolution of the Heisenberg spin system
and therefore obtained the formally exact solutions of the
Schr\"{o}dinger equation governing this spin system in a
time-dependent magnetic field\cite{Yan}. The relationship between
the unitary transformations of the dynamics of quantum systems
with time-dependent Hamiltonians and the gauge theories has been
found by Montesinos {\it et al.} \cite {Montesinos}. The magnetic
moment in time-dependent magnetic fields possesses so many
interesting and significant effects and properties that we think
that the dual of the magnetic moment, i. e., the electric dipole
moment in time-dependent exterior fields\cite {Wang} may also have
such similarly significant effects and properties. Note, however,
that the spinning magnetic moment of quantum systems is the
quantum mechanical effect and therefore the generators of the
Hamiltonian form the Lie algebraic commuting relations, which
enables us to easily solve the time-dependent Schr\"{o}dinger
equation governing the time evolution of these quantum spinning
systems, whereas the electric dipole is not the quantum mechanical
effect and therefore no such Lie algebraic commuting relations
exist in its Hamiltonian. Hence, we should quantize the electric
dipole and consider it a quantum dipole oscillator driven by an
exterior force, and then exactly solve the Schr\"{o}dinger
equation that governs this driven dipole oscillator in the
presence of the time-dependent exterior electric field.

Berry$^{,}$s discovery that the adiabatic geometric phase of wave
function arises in adiabatic quantum process\cite {Berry} opens up
new opportunities for investigating the global or topological
properties of quantum evolution\cite
{Furtado,Shen1,Kuppermann,Wagh,Sanders}. It is now well known that
the geometric phase appears in systems with the time-dependent
Hamiltonian, or in systems whose Hamiltonian possesses evolution
parameters\cite{Shen2,Shen3}. Differing from the dynamical phase
that depends on dynamical quantities of systems, the geometric
phase is independent of these dynamical quantities such as energy,
frequency, velocity as well as coupling coefficients. Instead, it
is only related to the geometric nature of the pathway along which
quantum systems evolve. This, therefore, implies that the
geometric phase presents the topological or global properties of
quantum systems in time-evolution process, and that it possesses
the physical significance and can thus be applied to various
fields of physics\cite{Gong,Taguchi,Falci}. Geometric phases
attract attentions of many physicists in many fields such as
gravity theory\cite {Furtado,Shen1}, differential
geometry\cite{Simon}, atomic and molecular physics\cite
{Kuppermann}, nuclear physics\cite {Wagh}, quantum
optics\cite{Gong,Zhu,Yang}, condensed matter
physics\cite{Taguchi,Falci,Exner}, molecular systems and chemical
reaction\cite {Kuppermann} as well.

Note, however, that Berry's adiabatic quantum theory can only deal
with the adiabatic evolution of quantum systems\cite {Berry}. If a
quantum system is in the presence of the fast varying exterior
field, then Berry's adiabatic theory is not appropriate to
investigate the non-adiabatic non-cyclic time evolution of wave
function of this quantum system.
In this situation we should obtain the exact solutions of the time-dependent Schr\"{o}%
dinger equation by using the invariant theory\cite {Lewis} and the
invariant-related unitary transformation formulation\cite {Gao1}.
It follows from the study of the spin model referred to earlier
that the resolution of the time-dependent Schr\"{o}%
dinger equation is often accompanied with obtaining the geometric
phase in the quantum evolution process. The invariant theory
suggested by Lewis and Riesenfeld\cite {Lewis} can exactly solve
the time-dependent Schr\"{o}dinger equation. Gao {\it et al.}
proposed the generalized invariant theory\cite{Gao1,Gao2}, by
introducing the basic invariants which enable one to find the
complete set of commuting invariants for some time-dependent
multi-dimensional systems\cite{Gao2}. Since the exact solutions
for systems with the time-dependent Hamiltonians obtained by the
invariant-related unitary transformation formulation contain both
geometric and dynamical phase and, fortunately, the result is
explicit rather than formal ( namely, there exists no
chronological product operator in wave function ), the
Lewis-Riesenfeld theory thus
developed into a powerful tool for treating the time-dependent Schr\"{o}%
dinger equation and the geometric phase factor.

Consider a dipole oscillator in the $x$-$y$ plane of the Cartesian
coordinate system with the time-dependent Hamiltonian
$H(t)=H_{1}(t)+H_{2}(t)+\frac{1}{2}(\omega _{1}+\omega _{2})$ ( in
the unit $\hbar =1$ ), where

\begin{equation}
H_{1}(t)=\omega _{1}a^{\dagger }a+c(t)(a^{\dagger }+a), \quad
H_{2}(t)=\omega _{2}b^{\dagger }b+s(t)(b^{\dagger }+b)   \eqnum{1}
\label{eq1}
\end{equation}
with the time-dependent parameters $c(t)=QE\sqrt{\frac{1}{2\mu
\omega _{1}}}\cos \Omega t,s(t)=QE\sqrt{\frac{1}{2\mu \omega
_{2}}}\sin \Omega t$. In the expression for the Hamiltonian,
$a^{\dagger }$ $(a)$ and $b^{\dagger }$ $(b)$ respectively stand
for the creation ( annihilation ) operators of the driven dipole
oscillators in $x$- and $y$- directions of the Cartesian
coordinate; $Q$ and $E$ respectively denote the polarization
charge of the electric dipole and the external electric field
strength, which are considered the time-independent parameters,
and $\Omega $ presents the rotating frequency of the exterior
rotating electric field. The exact solution of the time-dependent
Schr\"{o}dinger equation is $\left| \Psi (t)\right\rangle =\exp
[\frac{1}{2i}(\omega _{1}+\omega _{2})t]\left| \Psi
_{1}(t)\right\rangle \left| \Psi _{2}(t)\right\rangle $, where
$\left| \Psi _{1}(t)\right\rangle $ and $\left| \Psi
_{2}(t)\right\rangle $ respectively satisfy the following equation

\begin{equation}
H_{1}(t)\left| \Psi _{1}(t)\right\rangle =i\frac{\partial
}{\partial t}\left| \Psi _{1}(t)\right\rangle  \eqnum{2}
\label{eq2}
\end{equation}
and

\begin{equation}
H_{2}(t)\left| \Psi _{2}(t)\right\rangle =i\frac{\partial
}{\partial t}\left| \Psi _{2}(t)\right\rangle. \eqnum{3}
\label{eq3}
\end{equation}
We first solve Eq. (\ref{eq2}) as an illustrative example by
making use of the Lewis-Riesenfeld invariant theory and the
invariant-related unitary transformation formulation. In
accordance with the Lewis-Riesenfeld invariant theory, the
particular exact particular solution $\left| \Psi
_{1}(t)\right\rangle $ of the time-dependent Schr\"{o}dinger
equation Eq. (\ref{eq2}) is different from the eigenstate of the
invariant $I(t)$ only by a time-dependent $c$- number factor $\exp
\{\frac{1}{i}\int_{0}^{t}\left\langle \lambda ,t^{\prime }\right|
[H(t^{\prime })-i\frac{\partial }{\partial t^{\prime }}]\left|
\lambda ,t^{\prime }\right\rangle {\rm d}t^{\prime }\}$ with
$\left| \lambda ,t\right\rangle $ being the eigenstate of the
invariant $I(t)$ ( corresponding to the particular eigenvalue
$\lambda$ ) and satisfying the following eigenvalue equation

\begin{equation}
I(t)\left| \lambda ,t\right\rangle =\lambda \left| \lambda
,t\right\rangle,\eqnum{4}                  \label{eq4}
\end{equation}
where the eigenvalue $\lambda$ of the invariant $I(t)$ is
time-independent, so that it is the subject of this step to obtain
the eigenstate $\left| \lambda ,t\right\rangle$ of the invariant
$I(t)$. The meanings of the invariant $I(t)$ of the time-dependent
dipole oscillator is that $I(t)$ is a conserved operator and
agrees with the following Liouville-Von Neumann equation

\begin{equation}
\frac{\partial I(t)}{\partial t}+\frac{1}{i}[I(t),H_{1}(t)]=0.
\eqnum{5}                  \label{eq5}
\end{equation}
It follows from Eq. (\ref{eq5}) that the invariant $I(t)$ may be
constructed in terms of $a^{\dagger }a$,  $ a^{\dagger }$,  $ a$
and the time-dependent $c$- number parameter $\Delta (t)$ as
follows

\begin{equation}
I(t)=\alpha a^{\dagger }a+\eta (t)a^{\dagger }+\eta ^{\ast
}(t)a+\Delta (t),\eqnum{6}                  \label{eq6}
\end{equation}
where $\alpha$ is readily proved a time-independent $c$- number
coefficient. From the Liouville-Von Neumann equation (\ref{eq5}),
it follows that there exists a set of auxiliary equations

\begin{equation}
\dot{\eta}+\frac{1}{i}(\alpha c-\eta \omega _{1})=0, \quad
\dot{\eta}^{\ast }-\frac{1}{i}(\alpha c-\eta ^{\ast }\omega
_{1})=0, \quad \dot{\Delta}+\frac{1}{i}(\eta ^{\ast }-\eta
)c=0,\eqnum{7}                  \label{eq7}
\end{equation}
which is used to determine the time-dependent parameters, $\eta
(t)$, $\eta ^{\ast }(t)$, and $\Delta (t)$, of the invariant.

It should be noted that we cannot directly solve Eq. (\ref{eq4})
for the reason that the time-dependent parameters $\eta (t)$,
$\eta ^{\ast }(t)$, and $\Delta (t)$ are involved in Eq.
(\ref{eq6}). If, for example, we can find ( or construct ) a
unitary transformation operator $V(t)$ to make $V^{\dagger
}(t)I(t)V(t)$ be time-independent, then this problem of the
eigenvalue equation of the invariant $I(t)$ is therefore resolved.
According to our experience for utilizing the invariant-related
unitary transformation formulation, we suggest a following unitary
transformation operator

\begin{equation}
V(t)=\exp [-A(t)],\quad V^{\dagger }(t)=\exp [A(t)], \eqnum{8}
\label{eq8}
\end{equation}
where $A(t)=\beta (t)a^{\dagger }-\beta ^{\ast }(t)a$ with $\beta
(t)$ and $\beta ^{\ast }(t)$ being determined in what follows by
calculating $I_{V}=V^{\dagger }(t)I(t)V(t)$.

Calculation of $I_{V}=V^{\dagger }(t)I(t)V(t)$ yields

\begin{equation}
I_{V}=\alpha a^{\dagger }a+(\eta -\beta \alpha )a^{\dagger }+(\eta
^{\ast }-\beta ^{\ast }\alpha )a+(\Delta -\beta \eta ^{\ast
}-\beta ^{\ast }\eta +\beta ^{\ast }\beta \alpha ).\eqnum{9}
\label{eq9}
\end{equation}
If $\beta $ and $\beta ^{\ast }$ are chosen to be $\beta
=\frac{\eta }{\alpha },\beta ^{\ast }=\frac{\eta ^{\ast }}{\alpha
}$, then by using the auxiliary equations (\ref{eq7}) one can
arrive at

\begin{equation}
\Delta -\beta \eta ^{\ast }-\beta ^{\ast }\eta +\beta ^{\ast
}\beta \alpha =0.\eqnum{10}                  \label{eq10}
\end{equation}
This, therefore, means that we can change the time-dependent
$I(t)$ into a time-independent $I_{V}$, and the result is
$I_{V}=\alpha a^{\dagger }a$. The eigenvalue equation of $I_{V}$
is $I_{V}\left| n\right\rangle =n\alpha \left| n\right\rangle $,
and consequently the eigenvalue equation of $I(t)$ is
$I(t)V(t)\left| n\right\rangle=n\alpha V(t)\left| n\right\rangle$.
Further analysis shows that\cite {Lewis} the exact particular
solution $\left| \Psi _{1}(t)\right\rangle $ of the time-dependent
Schr\"{o}dinger equation Eq. (\ref{eq2}) is different from the
eigenstate of the invariant $I(t)$ only by a time-dependent $c$-
number factor $\exp [\frac{1}{i}\varphi (t)]=\exp
\{\frac{1}{i}\int_{0}^{t}\left\langle n \right| V^{\dagger
}(t^{\prime })[H(t^{\prime })-i\frac{\partial }{\partial t^{\prime
}}]V(t^{\prime })\left|n \right\rangle {\rm d}t^{\prime }\}$.

Correspondingly, $H_{1}(t)$ is transformed into

\begin{eqnarray}
H_{1V}(t) &=&V^{\dagger }(t)H_{1}(t)V(t)-V^{\dagger
}(t)i\frac{\partial }{\partial t}V(t)\nonumber \\
&=&\omega _{1}a^{\dagger }a+[c(t)-\beta (t)\omega
_{1}+i\dot{\beta}(t)]a^{\dagger }+[c(t)-\beta ^{\ast }(t)\omega
_{1}-i\dot{\beta}^{\ast }(t)]a\nonumber \\
&+&[\beta ^{\ast }(t)\beta (t)\omega _{1}-\beta (t)c(t)-\beta
^{\ast }(t)c(t)]+\frac{i}{2}[\dot{\beta}(t)\beta ^{\ast
}(t)-\dot{\beta}^{\ast }(t)\beta (t)]   \eqnum{11} \label{eq11}
\end{eqnarray}
under the unitary transformation $V(t)$. By using the auxiliary
equations (\ref{eq7}) and $\beta =\frac{\eta }{\alpha },\beta
^{\ast }=\frac{\eta ^{\ast }}{\alpha }$, it is verified that
$c(t)-\beta (t)\omega _{1}+i\dot{\beta}(t)=0$ and $c(t)-\beta
^{\ast }(t)\omega _{1}-i\dot{\beta}^{\ast }(t)=0$ in $H_{1V}(t)$.
Thus $H_{1V}(t)$ is rewritten as

\begin{equation}
H_{1V}(t) =\omega _{1}a^{\dagger }a+[\beta ^{\ast }(t)\beta
(t)\omega _{1}-\beta (t)c(t)-\beta ^{\ast
}(t)c(t)]+\frac{i}{2}[\dot{\beta}(t)\beta ^{\ast
}(t)-\dot{\beta}^{\ast }(t)\beta (t)]   \eqnum{12} \label{eq12}
\end{equation}
and the time-dependent $c$- number factor $\exp
[\frac{1}{i}\varphi (t)]$ is therefore

\begin{eqnarray}
&&\exp \{\frac{1}{i}\int_{0}^{t}\left\langle n \right| V^{\dagger
}(t^{\prime })[H(t^{\prime })-i\frac{\partial }{\partial t^{\prime
}}]V(t^{\prime })\left|n \right\rangle {\rm d}t^{\prime }\}\nonumber \\
&=&\exp \{\frac{1}{i}\int_{0}^{t}[n\omega _{1}+\beta ^{\ast
}(t^{\prime })\beta (t^{\prime })\omega _{1}-\beta (t^{\prime
})c(t^{\prime })-\beta ^{\ast }(t^{\prime })c(t^{\prime
})+\frac{i}{2}[\dot{\beta}(t^{\prime })\beta ^{\ast }(t^{\prime
})-\dot{\beta}^{\ast }(t^{\prime })\beta (t^{\prime })]]{\rm
d}t^{\prime }\}.\eqnum{13}                  \label{eq13}
\end{eqnarray}
Hence the particular exact solution of the time-dependent
Schr\"{o}dinger equation (\ref{eq2}) corresponding to the
particular eigenvalue, $n$, of the invariant $I(t)$ may be written

\begin{equation}
\left| \Psi _{1}(t)\right\rangle =\exp [\frac{1}{i}\varphi
(t)]V(t)\left| n\right\rangle .\eqnum{14} \label{eq14}
\end{equation}
In the same fashion, one can easily obtain the particular exact
solution of the time-dependent Schr\"{o}dinger equation
(\ref{eq3}) with which $\left| \Psi _{2}(t)\right\rangle $ agrees
and in consequence $\left| \Psi (t)\right\rangle $ is thus exactly
solved by using the Lewis-Riesenfeld invariant theory.

We obtain

\begin{eqnarray}
\eta (t)&=&B\exp (-i\omega _{1}t)-\frac{\alpha
QE\sqrt{\frac{1}{2\mu \omega _{1}}}(-\omega _{1}\cos \Omega
t+i\Omega \sin \Omega t)}{\omega _{1}^{2}-\Omega ^{2}},\nonumber \\
\eta ^{\ast }(t)&=&B\exp (i\omega _{1}t)-\frac{\alpha
QE\sqrt{\frac{1}{2\mu \omega _{1}}}(-\omega _{1}\cos \Omega
t-i\Omega \sin \Omega t)}{\omega _{1}^{2}-\Omega ^{2}}\eqnum{15}
\label{eq15}
\end{eqnarray}
from the auxiliary equations $\dot{\eta}+\frac{1}{i}(\alpha c-\eta
\omega _{1})=0$ and $\dot{\eta}^{\ast }-\frac{1}{i}(\alpha c-\eta
^{\ast }\omega _{1})=0$. Thus the time-dependent parameters $\beta
(t) $ and its complex conjugate $\beta ^{\ast }(t)$ in the
particular exact solutions of the time-dependent Schr\"{o}dinger
equations are solved by using the relations $\beta (t) =\frac{\eta
(t)}{\alpha }$ and $\beta ^{\ast }(t)=\frac{\eta ^{\ast
}(t)}{\alpha }$. If we choose the coefficient $B$ to be
$-\frac{\alpha QE\sqrt{\frac{\omega _{1}}{2\mu }}}{\omega
_{1}^{2}-\Omega ^{2}}$, then the time-evolution operator,
$U_{1}(t)=\exp [\frac{1}{i}\varphi (t)]V(t)$, of $\left| \Psi
_{1}(t)\right\rangle $ may be $U_{1}=1$ at the initial $t=0$.
Apparently, it is seen as declared above that the exact solution
obtained here does not involve the chronological product operator
and may be considered the explicit solution.

The global property of the geometric phase $\varphi
^{(g)}(t)=\frac{i}{2}\int_{0}^{t}\{[\dot{\beta}(t^{\prime })\beta
^{\ast }(t^{\prime })-\dot{\beta}^{\ast }(t^{\prime })\beta
(t^{\prime })]\}{\rm d}t^{\prime }$ may be seen from the following
flux expression for

\begin{equation}
\varphi ^{(g)}(T)=\frac{i}{2}\oint_{C}[(\nabla \beta )\beta ^{\ast
}-(\nabla \beta ^{\ast })\beta ]\cdot {\rm d}{\bf
l}=i\int\int_{{\bf{s}}}[(\nabla \beta ^{\ast })\times (\nabla
\beta )]\cdot {\rm d}{\bf{s}},\eqnum{16} \label{eq16}
\end{equation}
with the calculation being performed in the parameter space where
the nabla $\nabla $ is defined to be ${\rm d}t={\rm d}{\bf l}\cdot
\nabla $. The global and topological property of the geometric
phase $\varphi
^{(g)}(t)=\frac{i}{2}\int_{0}^{t}\{[\dot{\beta}(t^{\prime })\beta
^{\ast }(t^{\prime })-\dot{\beta}^{\ast }(t^{\prime })\beta
(t^{\prime })]\}{\rm d}t^{\prime }$ may also be shown by comparing
it with the connection $1$- form $i\left\langle z\right| {\rm
d}\left| z\right\rangle =-{\rm Im}z^{\ast}z=\frac{1}{2}(p{\rm
d}q-q{\rm d}p)$ of the Glauber's coherent state $\left|
z\right\rangle =\exp (-\frac{1}{2}z^{\ast }z)\sum_{n=0}^{\infty
}\frac{z^{n}}{\sqrt{n!}}\left| n\right\rangle $ where
$z=\frac{1}{\sqrt{2}}(q+ip)$. Since it is well known that the loop
integration $\gamma =\frac{1}{2}\oint_{C}i\left\langle z\right|
{\rm d}\left| z\right\rangle $ with the curve $C$ being in the
complex $z$- phase plane presents the area surrounded by the curve
in the complex $z$- plane and is considered the Berry's phase, and
the expression for geometric phase $\varphi ^{(g)}(t)$ is in
analogy with $\gamma $, the topological property of the geometric
phase $\varphi ^{(g)}(t)$ is thus indicated by this comparison. It
is worth noticing again that the Berry's quantum phase theory\cite
{Berry} can only deal with the adiabatic case of the above quantum
system ( i. e., where the rotating frequency $\Omega $ of the
varying exterior field is very small ). It is readily verified
that when $\Omega \rightarrow 0$, the geometric phase expressed by
(\ref{eq16}) is reduced to the Berry's phase that can be
calculated by the Berry's phase formula\cite {Berry}. In our
formulation, however, we can study the non-adiabatic non-cyclic
geometric phase of the dipole oscillator, even the geometric phase
in the case of $\Omega \rightarrow \infty $.

This letter considers the exact solution and the geometric phase
of the electric dipole in the presence of the swiftly varying
electric field. Since the Berry's adiabatic quantum theory can
only deal with the adiabatic geometric phase and the
time-dependent evolution of dipole oscillator in the slowly
varying electric field, we make use of the Lewis-Riesenfeld theory
and obtain the exact solution of the time-dependent
Schr\"{o}dinger equation governing the time evolution of the
dipole oscillator and the expression for the geometric phase
arising in the non-adiabatic non-cyclic evolution process caused
by the exterior field. Since we have obtained the exact solution
of the time-dependent Schr\"{o}dinger equation and in concequence
get the expression for the non-adiabatic geometric phase of the
dipole oscillator in varying environment, and all these results
are explicit rather than formal ( i. e., no chronological product
operator is involved in the exact solution and the geometric phase
), we think that the method presented in this letter has several
advantages over those based on the Berry's adiabatic phase formula
and deserves further consideration.

Acknowledgements  The author greatly acknowledges helpful
suggestions from X. C. Gao.

\end{document}